\documentclass[10pt,twocolumn,reqno, aps,prb,superscriptaddress]{revtex4-2}
\usepackage{amsmath}
\usepackage[pdftex]{graphicx}
\usepackage{float}
\usepackage{color}
\usepackage{rotating}
\usepackage[breaklinks=true,colorlinks,citecolor=blue,linkcolor=blue,urlcolor=blue]{hyperref}
\usepackage{graphicx}
\usepackage{natbib}
\usepackage{dcolumn}
\usepackage{bm}
\usepackage{hyperref}
\usepackage[caption=false]{subfig}
\usepackage{epstopdf}
\usepackage{ulem}
\usepackage{amsfonts}
\usepackage{amssymb}

\providecommand{\U}[1]{\protect\rule{.1in}{.1in}}
\begin{document}

\title{Cryogenic spin Seebeck effect}
\author{Mehrdad Elyasi}
\affiliation{Institute for Materials Research, Tohoku University, 2-1-1 Katahira, 980-8577 Sendai, Japan
}
\author{Gerrit E. W. Bauer}
\affiliation{Institute for Materials Research \& AIMR \& CSRN, Tohoku
University, 2-1-1 Katahira, 980-8577 Sendai, Japan} 
\affiliation{Zernike
Institute for Advanced Materials, University of Groningen, The Netherlands}
\date{\today }

\begin{abstract}
We present a theory of the non-linearities of the spin Seebeck effect (SSE)
in a ferromagnetic nanowire at cryogenic temperatures. We adopt a
microscopic quantum noise model based on a collection of two-level systems.
At certain positions of Pt detectors to the wire, the transverse SSE changes
sign as a function of temperature and/or temperature gradient. On the other
hand, the longitudinal SSE does not show significant non-linearities even far
outside the regime of validity of linear response theory.
\end{abstract}

\maketitle

We address the spin Seebeck effect (SSE) in electrically insulating magnets,
i.e. the spin current caused by a temperature gradient as detected by the
inverse spin Hall effect voltage in heavy metal contacts \cite%
{Uchida2010,Xiao2010,Adachi2011,Bauer2012,Ohnuma2013,Wu2015,Wu2016}. The
longitudinal SSE (LSSE) is observed in a planar configuration in which the
heat and spin currents flow in parallel and normal to the interfaces \cite%
{Uchida2010b}. The transverse or non-local \cite{Cornelissen2015} SSE (TSSE)
refers to more complicated configurations, usually two contacts on the
surface of a magnetic slab or film. The spin current is injected into the
metal contact by spin pumping \cite{Xiao2010} but the signal is usually
dominated by the currents that are generated by temperature gradients in the
bulk of the magnet \cite{Rezende2014}. The reported signals are in general
proportional to the applied temperature differences $\triangle T.$ However,
several recent studies of the SSE at low temperatures \cite%
{Kikkawa2015,Jin2015,Schreier2016,Guo2016,Iguchi2017,De2020,Oyanagi2020,Ganzhorn2020}
do not address a fundamental issue of thermal transport at ultralow
temperatures. Linear response is valid when the perturbation is sufficiently
small, but the properly normalized driving force is not $\triangle T$ but $%
\triangle T/T$ (or $\partial T/T),$ i.e. the temperature difference divided
by the average one \cite{Onsager}. This condition is increasingly difficult
to fulfill at low temperatures, or positively formulated, it should become
easier to access non-linear thermomagnonic transport phenomena.

Existing theoretical treatments of the spin Seebeck effect are not suitable
to address the low-temperature and non-linear regimes. The low frequency
magnons that dominate at cryogenic temperatures are strongly affected by
dipolar interactions, so exchange-only magnon models fail. The assumption of
a semiclassical magnon accumulation in terms of a local chemical potential
and magnon temperature \cite{Cornelissen2016} breaks down because
thermalization becomes weak. With a classical magnetization noise model and
in linear response, the non-thermal distribution functions governing the SSE
can be described by mode- (rather than position-) dependent magnon
temperatures and chemical potentials \cite{Yan2017}. Treatments of the
stochastic magnetization dynamics in terms of classical white noise sources 
\cite{Ohe2011,Chotorlishvili2013,Ritzmann2015,Barker2016} do not work at low
temperatures. This can be repaired by a noise spectrum that obeys the
quantum fluctuation dissipation theorem \cite{Barker2019}, but at the cost
of introducing phenomenological damping constants. A recent linear response
study of the LSSE at low temperatures \cite{Schmidt2020} focuses on the
magnon-polaron hybrid state at large magnetic fields \cite{Kikkawa2016}.

The broadening of the ferromagnetic resonance of a YIG sphere increases $%
\propto T$ for $T>1\,$K. The minimum in the damping followed by an increase
and saturation with decreasing temperatures\ $<1\,$K \cite{Tabuchi2014} is
caused by impurities and disorder, presumably two-level systems (TLS) \cite%
{Spencer1961,Vanvleck1963,Vanvleck1964}. The spin and heat transport in this
regime has to our knowledge not been addressed in the literature and is the
focus of this Letter. We study the cryogenic SSE of a ferromagnetic (FM)
nanowire with a microscopic TLS\ model for the thermal noise at weak
magnetic fields. In this regime magnon-magnon and magnon-phonon interactions
may be safely disregarded. We predict that the antisymmetry of the TSSE
signal as a function of position of a Pt detector \cite%
{Xiao2010,Adachi2011,Ohe2011} is broken in the non-linear regime and a
non-monotonous temperature dependence at certain contact positions emerges.
These effects are caused by the non-uniform gradient of the spin
distribution functions in spite of a constant temperature gradient. The LSSE
signal is on the other hand surprisingly robust, with a linear dependence on
a global temperature difference $\Delta T$ much larger than $T_{M}$.

\textit{Model}. We consider YIG nanostructures with high quality surfaces 
\cite{GSchmidt2020} in which scattering at low temperatures is dominated by
rare earth (RE) substitutional impurities, e.g. Tb or Yb, on the Y sites 
\cite{Tabuchi2014,Spencer1961,Vanvleck1963,Vanvleck1964}. Two degenerate
atomic levels of a RE atom form a two-level system (TLS) with pseudo-spin $%
\vec{\Omega}$ that interacts with the local iron magnetic moments of spin $%
\vec{S}_{Fe}$ by an exchange interaction $H_{TLS}=\vec{S}_{Fe}\cdot \bar{K}%
\vec{\Omega}$, where $\bar{K}$ is an anisotropic exchange interaction
tensor, which splits the pseudo-spin levels by $\omega _{01}$. Since the RE
angular momentum strongly couples to the lattice, spin waves can be
efficiently dissipated via $H_{TLS}$. The isotropic Heisenberg exchange
contribution $S_{x(y)}\Omega _{x(y)}$ couples the precessional dynamics and
leads to a \textquotedblleft transverse\textquotedblright\ relaxation that
preserves the total magnetization. The anisotropy introduces
\textquotedblleft longitudinal\textquotedblright\ terms like $S_{x}\Omega
_{z}$ and $S_{y}\Omega _{z}$ by which the splitting $\omega _{01}$ depends
on the magnetization direction. Van Vleck \cite{Vanvleck1963,Vanvleck1964}
computed the life time broadenings due to $H_{TLS}$ as a function of the
ratio of rare earth to Fe concentration $c$. For the longitudinal\ process
he reports 
\begin{figure}[t]
\includegraphics[width=0.5\textwidth]{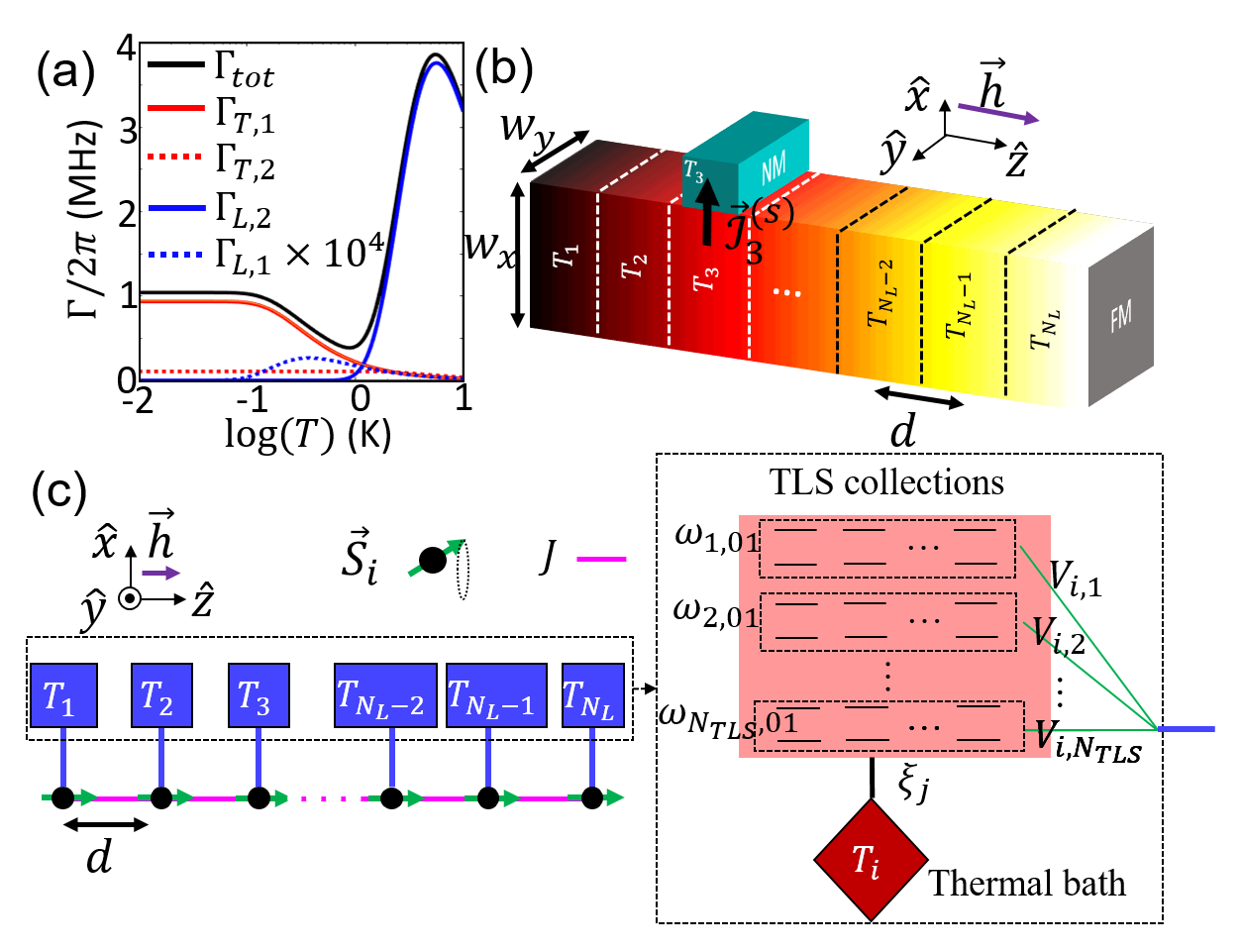}
\caption{Model. (a) The dissipation caused by two TLS ensembles that fit the
experiments of Tabuchi et al. \protect\cite{Tabuchi2014}. $\Gamma
_{tot}=\sum_{j=1,2}\Gamma _{L(T),j}$, where $j$ indicates the ensembles
parametrized by $c_{1}=3\times 10^{-7}$, $c_{2}=1\times 10^{-4}$, $\protect%
\tau _{1}=10\,$ns, $\protect\tau _{2}=0.1\,$ps, $\protect\omega _{1,01}=2%
\protect\pi \times 10\,$GHz, $\protect\omega _{2,01}=2\protect\pi \times
150\,$GHz and $\protect\omega _{U}=2\protect\pi \times 10\,$GHz. (b) A spin
Seebeck current $\mathcal{J}^{(s)}$ polarized along $\hat{z}$ flows from the
magnet into the metal contact. The color indicates the temperature profile,
where white (black) is hottest (coldest). (c) Left: Array of spins $\vec{S}%
_{i}$, coordinate system, lattice spacing $d$, external magnetic field $\vec{%
h}$, and local reservoirs at temperature $T_{i}$. Right:
mesoreservoir of $N_{TLS}$ two-level systems (TLS) with frequency splittings 
$\protect\omega _{j,01}$. The TLS ensemble is in contact with a thermal bath
(relaxation rate $\protect\xi _{j}=2\protect\pi /\protect\tau _{j}$) at
temperature $T_{i}$ and interacts with a spin (green lines) by $V_{i,j}$.}
\label{fig1}
\end{figure}
\begin{align}
\Gamma _{L}& =\sum_{j}\Gamma _{L,j}=\sum_{j}\frac{c_{j}\hbar }{6k_{B}T}%
\sum_{n=1}^{6}  \notag \\
& \omega _{j,n,01}^{2}f_{j,n}(\theta ,\phi )\frac{\tau _{j,n}\omega _{U}}{%
1+\tau _{j,n}^{2}\omega _{U}^{2}}\left[ 1-\tanh ^{2}\frac{\hbar \omega
_{j,n,01}}{2k_{B}T}\right] ,  \label{eq1}
\end{align}%
where $j$ indicates a certain type of impurity or a certain corresponding
TLS, while $n$ indicates a yttrium site in the YIG unit cell. $\tau _{j,n}$
is the relaxation time of an excited TLS, $f_{j,n}(\theta ,\phi )\in \left[
0,1\right] $ depends on the polar magnetization direction angles $\theta
,\phi ,$ and $\omega _{U}$ is the FMR frequency. The transverse relaxation
is dominated by the isotropic exchange and reads \cite%
{Vanvleck1963,Vanvleck1964} 
\begin{align}
\Gamma _{T}& =\sum_{j}\Gamma _{T,j}=-\mathrm{Im}\sum_{j}\sum_{n=1}^{6}\frac{%
c_{j}\omega _{j,n,01}}{12}\tanh \frac{\hbar \omega _{j,n,01}}{2k_{B}T}\times
\notag \\
& \left[ \frac{\omega _{U}}{\omega _{U}-\omega _{j,n,01}+i/\tau _{j,n}}+%
\frac{\omega _{U}}{\omega _{U}+\omega _{j,n,01}+i/\tau _{j,n}}\right] .
\label{eq2}
\end{align}%
For $\omega _{j,n,01}=\omega _{j,01}\forall n$ and $\tau _{j,n}=\tau
_{j}\forall n$, 
\begin{align}
\Gamma _{T,j}& =\frac{1}{2}\omega _{U}c_{j}\tanh \frac{\hbar \omega _{j,01}}{%
2k_{B}T}\times  \notag \\
& \left\{ 
\begin{array}{c}
\omega _{j,01}\tau _{j}/\left( 1+\tau _{j}^{2}\omega _{j,01}^{2}\right) \\ 
\omega _{j,01}\tau _{j}%
\end{array}%
\ \mathrm{for}\ 
\begin{array}{c}
\omega _{j,01}\gg \omega _{U} \\ 
\omega _{j,01}\sim \omega _{U}%
\end{array}%
.\right.  \label{eq3}
\end{align}%
$\Gamma _{L,j}$ is a non-monotonous function of temperature, increasing from
zero at $T=0\ $up to a maximum at $\sim \hbar \omega _{01}/k_{B}$. $\Gamma
_{T,j}$ monotonically increases from zero as $T$ decreases and saturates to
a finite value at $T=0$ since the transverse relaxation is proportional to
polarization of the TLS, i.e. $\tanh \left[ \hbar \omega _{j,01}/\left(
2k_{B}T\right) \right] /2$. The proportionality of $\Gamma _{T,j}$ with $\tau
_{j}$ when $\omega _{j,01}\sim \omega _{U}$, only holds for $1/\tau _{j}\ll
\omega _{U}.$ $\Gamma _{T,j}$ vanishes with $\tau _{j}$ because of the
associated lifetime broadening of the TLS density of states. Tabuchi et al. 
\cite{Tabuchi2014} found an excellent agreement for the temperature
dependent broadening at $T<1\,$K assuming $\omega _{01}/2\pi \sim \omega
_{U}/2\pi \sim 10\,\mathrm{GHz}$, and a temperature independent bias. The
increase in damping for $T\geq 1\,$K might be phonon induced, but could
indicate also the existence of a second family of levels with larger
exchange splitting. Figure \ref{fig1}(a) shows that the total dissipation
due to combination of two distinct TLS, $\Gamma _{tot}=\sum_{j=1,2}\left(
\Gamma _{L,j}+\Gamma _{T,j}\right) $ explains the observed damping very well
for up to $T\sim 5\,$K, where we used $c_{1}=3\times 10^{-7}$, $%
c_{2}=1\times 10^{-4}$, $\tau _{1}=10\,$ns, $\tau _{2}=10\,$ps, $f_{1}=1$, $%
f_{2}=1$, $\omega _{1,01}/2\pi =10\,$GHz, and $\omega _{2,01}/2\pi =150\,$%
GHz. Figure \ref{fig1}(a) shows that at $T<1\,$K, $\Gamma _{T,1}\gg \Gamma
_{L,1(2)},\Gamma _{T,2}$. In the following we therefore consider only a
single TLS type with $\Gamma _{tot}\approx \alpha _{\mathrm{TLS}}\omega _{%
\text{U}}$, where $\alpha _{\mathrm{TLS}}=c_{1}\tau _{1}\omega _{1,01}\tanh
\left( \hbar \omega _{1,01}/2k_{B}T\right) \approx 10^{-4}$ is the (Gilbert)
damping coefficient. We proceed to predict the consequence of TLS dominated
dissipation for the spin Seebeck effect.

Figure \ref{fig1}(b) shows the schematics of the physical system, a nanowire
magnetized along its length, while Fig. \ref{fig1}(c) shows the schematics
of the corresponding spin lattice-reservoir model. The dipolar interactions
affect the magnon dispersion only for wave lengths that are much larger than
the unit cell. We therefore adopt a micromagnetic approach in which the
local magnetization represents an average over slices of typically 50 nm
that contain many local moments. The macrospin site $i$ then interacts with
a \textquotedblleft mesoreservoir\textquotedblright\ composed of several TLS
as described earlier. Since the latter are local impurities with short-range
exchange interactions, we may disregard their cross-correlation. The
mesoreservoir in turn interacts with a large reservoir with well-defined
temperature $T_{i}$ that is allowed to vary slowly in space. The Hamiltonian
for the model in Fig. \ref{fig1}(c) now reads 
\begin{equation}
H=H_{S}+H_{R}+H_{SR},  \label{eq4}
\end{equation}%
where $H_{S}$ describes the magnet, $H_{R}$ the mesoreservoirs, and $H_{SR}$
the interaction between them. We expand the Heisenberg Hamiltonian for the
spin chain of Fig. \ref{fig1}(c) to the second order of the
Holstein-Primakoff transformation, for a spin $S$ on site $i$, i.e. $%
S_{i}^{+}=\sqrt{2S}a_{i}^{\dag }{[1-a_{i}^{\dag }a_{i}/}\left( {2S}\right) {]%
}^{1/2}$, $S_{i}^{-}=\sqrt{2S}{[1-a_{i}^{\dag }a_{i}/}\left( {2S}\right) {]}%
^{1/2}a_{i}$, $S_{i}^{z}=S-a_{i}^{\dag }a_{i}$, in terms of magnons $%
a_{i}^{\dag }$($a_{i}$) created (annihilated) at site $i$. This leads to 
\begin{align}
H_{S}& =\sum_{i}\left( \mathcal{A}_{i}-S\sum_{j}F_{zz}^{i,j}\right)
a_{i}^{\dag }a_{i}+  \notag \\
& \sum_{i,j}\left\{ \left[ SJ\delta \left( i\pm 1,j\right) +\mathcal{B}%
_{i,j})\right] a_{i}^{\dag }a_{j}+\right.  \notag \\
& \left. \mathcal{C}_{i,j}a_{i}a_{j}+\mathrm{H.C.}\right\} ,  \label{eq5}
\end{align}%
where $\mathcal{A}_{i}=-2SJ+\gamma _{e}h^{z}$. $\mathcal{A}%
_{1(N_{L})}=-2SJ+\gamma _{e}h^{z}$ indicates that the edges are in contact
with a pinned spin, otherwise $\mathcal{A}_{1(N_{L})}=-SJ+\gamma _{e}h^{z}$. 
$h^{z}$ is the magnetic field in the $\hat{z}$ direction, $\gamma _{e}$ is
the gyromagnetic ratio, $\delta $ is the Kronecker delta, $\mathcal{B}%
_{i,j}=S\left( F_{xx}^{i,j}+F_{yy}^{i,j}\right) /2$, $\mathcal{C}%
_{i,j}=S\left( F_{xx}^{i,j}-F_{yy}^{i,j}\right) /2$, where $F_{xx(yy)}^{i,j}$
is the dipolar field of $S_{i}^{x}$ ($S_{i}^{y}$) exerted on $S_{j}^{x}$ ($%
S_{j}^{y}$), and $J$ is the exchange interaction\textit{.} We compute the
dipolar interactions assuming uniform dynamics along the thickness of the
nanowire ($\Vert \hat{x}$) and a nodeless cosine function amplitude with an
effective width along $\hat{y}$ \cite{Wang2019}.

\begin{figure}[t]
\includegraphics[width=0.5\textwidth]{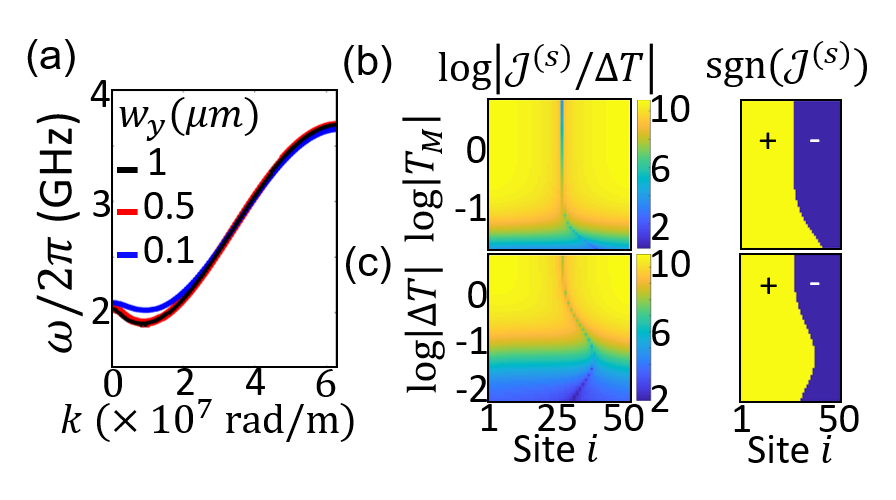}
\caption{(a) The magnon dispersion of a wire with different widths $w_{y}$.
The wavenumber $k$ corresponds to the peak of the Fourier transform of the
spatial wave function. (b) and (c) Spin Seebeck current as a function of
average temperature $T_{M}=(T_{L}+T_{R})/2$ and temperature difference $%
\Delta T=T_{R}-T_{L}$, respectively. Left: $|\mathcal{J}^{(s)}/\Delta T|$.
Right: $\mathcal{J}^{(s)}/|\mathcal{J}^{(s)}|$. In (b), $\Delta T=10\,$mK.
In (c), $T_{L}=20\,$mK. In (b) and (c), $w_{y}=500\,$nm [red curve in (a)].
In (a)-(c), $h^{z}=20\,$mT.}
\label{fig2}
\end{figure}

We parametrize $J$ by its value in the continuum limit. For long wave
lengths $J=\gamma _{e}\mu _{0}M_{s}\lambda ^{2}/\left( d^{2}S\right) ,$
where $S=NS_{0}$ and $N=w_{x}w_{y}d/l^{3}$ is the number of unit cells in
each 1D segment, $w_{x}$ ($w_{y}$) is the thickness (width) of the nanowire, 
$l$ is the unit cell dimension. $S_{0}=l^{3}M_{s}/\left( 2\pi \gamma
_{e}\right) \approx 14$ is the net number of spins in the YIG unit cell,
with magnetization $M_{s}=1.46\times 10^{5}\,$A/m, $\gamma _{e}=26\,$GHz/T
and $l=1.2\,$nm. The exchange length for YIG $\lambda =\sqrt{3}\times
10^{-8}\,\text{m}$ \cite{Stancil2009}. Figure \ref{fig2}(a) shows the
dipolar-exchange magnon dispersion for three values of $w_{y}$, where $%
d=50\, $nm, number of segments $N_{L}=200$, i.e. a nanowire of length $L=10\,%
\mathrm{\mu }$m. Figure \ref{fig2}(a) shows that the dispersion minimum
becomes deeper for larger $w_{y}$. When the wire is not too narrow (e.g. $%
w_{y}>100\,$nm for $w_{x}=100\,$nm \cite{Wang2019}), the dispersion relation
is non-monotonous or \textquotedblleft backward moving\textquotedblright\
for small wave vectors along the magnetization \cite%
{Kalinikos1986,Hurben1995,Elyasi2020}\textit{.}

Next, we Bosonize the Hamiltonian $H_{R}=\sum_{i}\sum_{j}\omega
_{j,01}r_{i,j}^{\dag }r_{i,j}$ and its interaction with the system $%
H_{SR}=\sum_{i}\sum_{j}\left( V_{i,j}r_{i,j}^{\dag }a_{i}+\mathrm{H.c.}%
\right) $, where $i$ labels the magnetic segments and $j$ the TLS for weak
excitations, i.e. $\hbar \omega _{j,01}\gg k_{B}T$. The TLS pseudo-spin $%
\Omega $ Hamiltonian can be simplified by another Holstein-Primakoff
transformation $\mathcal{L}_{i,j}^{+}=\sqrt{2\mathcal{L}_{j}}r_{i,j}^{\dag }$
and $\mathcal{L}_{i,j}^{-}=\sqrt{2\mathcal{L}_{j}}r_{i,j}$, where $%
r_{i,j}^{\dag }$ ($r_{i,j}$) creates (annihilates) a boson with frequency $%
\omega _{j,01}$. The polarization of a TLS with index $j$ in the collection, 
$\mathcal{L}_{j}=\langle \Omega _{z}\rangle =\tanh \left[ \hbar \omega
_{j,01}/\left( 2k_{B}T\right) \right] /2$. $V_{i,j}=\omega _{j,01}\sqrt{c_{j}%
\mathcal{L}_{j}}/\sqrt{S_{0}}$ is the interaction between a magnon on site $%
i $ with pseudo-spin $j$ at relative concentration $c_{j}$. Each TLS
collection is in contact with a large reservoir at a (slowly varying)
temperature $T_{i}$ and dissipation $\xi _{j}=2\pi /\tau _{j}$. This
dissipation is accompanied by the fluctuating field acting on TLS collection 
$g_{i,j}=\sqrt{\xi _{j}}\mathcal{F}_{i,j}$, where $\langle \mathcal{F}%
_{i,j}(t)\mathcal{F}_{i,j}^{\dag }(t^{\prime })\rangle =\left(
n_{i,j}^{th}+1\right) \delta (t-t^{\prime })$, $\langle \mathcal{F}%
_{i,j}^{\dag }\mathcal{F}_{i,j}\rangle =n_{i,j}^{th}\delta (t-t^{\prime })$,
and $n_{i,j}^{th}=\left( e^{\hbar \omega _{j,01}/k_{B}T_{i}}-1\right) ^{-1}$%
. The white noise correlation functions hold as long as $\hbar \xi _{j}\ll
k_{B}T_{i}$ for each $i$, which is a safe assumption for $T_{i}>10\,$mK and $%
\tau _{1}=10\,$ns. Here, we focus on low temperatures $T<1\,$K and a single
TLS parametrized by $c_{1}=3\times 10^{-7}$, $\tau _{1}=10\,$ns, and $\omega
_{1,01}/2\pi =10\,$GHz, leading to $V_{i,1}/2\pi \approx 1.5\,$MHz at $T=0$.
[see Fig. \ref{fig1}(a)]. We are safely in the regime $\langle r_{i,1}^{\dag
}r_{i,1}\rangle \ll Nc_{1}\mathcal{L}_{1}$, where $N$ is number of unit
cells, i.e. far from the saturation of TLS excitations.

We now address the steady state for the model defined above, i.e. a closed
system of a magnetic nanowire with a large temperature gradient and at low
temperatures. The Pt side contacts non-invasively detect the non-thermal
component of site-dependent magnon distributions, i.e. the TSSE, which we
compute numerically without additional approximations. The objective is the
matrix $\Lambda _{\infty }$ of the equal time correlation function of the
phase space variables $m_{x,i}=a_{i}+a_{i}^{\dag }$, $%
m_{y,i}=-i(a_{i}-a_{i}^{\dag })$, $X_{i}=r_{i}+r_{i}^{\dag }$, and $%
Y_{i}=-i(r_{i}-r_{i}^{\dag })$ (or symmetric covariance matrix) in the
steady state that governs the spatially dependent magnon population and spin
currents [see e.g. Eq. (\ref{eq8})] \cite{Carmichael1999}. This is the
long-time limit of the time-dependent covariance matrix ${\Lambda }$ that
obeys the equation of motion $\dot{\Lambda}=\mathcal{O}\Lambda +\Lambda 
\mathcal{O}+\Upsilon $ \cite{Carmichael1999}, where $\dot{\mathbf{v}}=%
\mathcal{O}\mathbf{v}+\mathbf{\mathfrak{c}}$, $\mathbf{v}=\left[
m_{x,1},m_{y,1},X_{1},Y_{1},\cdots ,m_{x,L},m_{y,L},X_{L},Y_{L}\right] $,
and $\mathcal{O}$ is determined by the Heisenberg equation $\mathbf{v}(p)=-i%
\left[ H,\mathbf{v}(p)\right] -\mathbf{\zeta }(p)\mathbf{v}(p)/2$. $\mathbf{%
\zeta }(p)=\xi _{1}$ for $p\in \{4(i-1)+3,4(i-1)+4\}$ $\forall i$, while $%
\mathbf{\zeta }(p)=0$ for $p\in \{4(i-1)+1,4(i-1)+2\}$ $\forall i$. $\mathbf{%
c}$ is the vector of fluctuating fields and determines $\Upsilon =\langle
\left( \mathbf{c}^{T}\mathbf{c}+\mathbf{c}\mathbf{c}^{T}\right) /2\rangle $. 
$\Upsilon $ is diagonal with elements $\Upsilon (p,p)=\zeta
(p)(2n_{i}^{th}+1)=\zeta (p)\left[ 2\left( e^{\hbar \omega
_{1,01}/k_{B}T_{i}}-1\right) ^{-1}+1\right] $ ($n_{i}^{th}$ is the Planck
distribution). We obtain $\Lambda _{\infty }$ by solving $\mathcal{O}\Lambda
_{\infty }+\Lambda _{\infty }\mathcal{O}=-\Upsilon $. The latter equation
can be cast into a linear system of equations in the phase space variables
that we solve numerically by inverting a (non-sparse) $(4\times
N_{L})^{2}\times (4\times N_{L})^{2}$ matrix, which in practice limits the
system size to $N_{L}<100$.

The spin Seebeck spin current can be detected by the inverse spin-Hall
voltage in Pt contacts generated by\ the spin current pumped by a
non-equilibrium magnetization at the YIG$|$Pt interface \cite%
{Tserkovnyak2002,Zhang2004,Xiao2010,Adachi2011}. The spin pumping at site $i$
\begin{equation}
\tilde{\mathcal{J}}_{i}^{(SP)}\approx \frac{\hbar g_{r}}{4\pi }\langle \vec{S%
}_{i}\times \dot{\vec{S}}_{i}\rangle =\frac{\hbar g_{r}}{4\pi }\langle
S_{i}^{x}\dot{S}_{i}^{y}-S_{i}^{y}\dot{S}_{i}^{x}\rangle ,
\end{equation}%
where $g_{r}$ is the real part of the complex spin mixing conductance. We
adress ${\mathcal{J}}_{i}^{(SP)}=4\pi \tilde{\mathcal{J}}_{i}^{(SP)}/\hbar
g_{r}$ in the rest of the paper. $\dot{S}^{x,(y,z)}=-i\left[ S^{x(y,z)},H_{S}%
\right] $ leads to 
\begin{align}
\mathcal{J}_{i}^{(SP)}& =\frac{1}{4}\sum_{j}\left\{ \left[ SJ\delta \left(
i\pm 1,j\right) +\mathcal{B}_{i,j}\right] \times \right.  \notag \\
& \left. \left( \langle m_{x}^{i}m_{x}^{j}\rangle +\langle
m_{y}^{i}m_{y}^{j}\rangle \right) +\mathcal{C}_{ij}\left( -\langle
m_{x}^{i}m_{x}^{j}\rangle +\langle m_{y}^{i}m_{y}^{j}\rangle \right)
\right\} ,  \label{eq8}
\end{align}%
which at equilibrium is canceled exactly by the torque induced by the
thermal spin current noise emitted by the metal contact \cite{Xiao2010}.
Therefore, for a certain temperature profile $T_i$, the net spin current
pumped from site $i$ into NM, $\mathcal{J}_{i}^{(s)}=\mathcal{J}%
_{i}^{(SP,NEQ)}-\mathcal{J}_{i}^{(SP,EQ)}$, where $\mathcal{J}%
_{i}^{(SP,NEQ)} $ ($\mathcal{J}_{i}^{(SP,EQ)})$ are the non-equilibrium
(equilibrium) currents at a contact $i$ with temperature $T_i$.
Disregarding any spin accumulation in the metal contacts, the SSE spin
current is pumped by non-equilibrium magnons. Indeed, the dominant term
(confirmed by calculations) in $J_{i}^{(SP)}$ is proportional to $\langle
m_{x}^{i}m_{x}^{i}\rangle +\langle m_{y}^{i}m_{y}^{i}\rangle =4\langle
a_{i}^{\dag }a_{i}\rangle +2$. Therefore, the local magnon accumulation at
each site, i.e. the difference of $\langle a_{i}^{\dag }a_{i}\rangle $ at
equilibrium and non-equilibrium, drives the spin current $\mathcal{J}%
_{i}^{(s)}$. The thermalization is weak so the local distribution functions
cannot be parametrized by magnon temperatures or chemical potentials. We
disregard the effect of the pumping on the magnon system for simplicity,
which is allowed when the mixing conductance is small, e.g. for sufficiently
small contacts. The distributed spin pumping currents in Eq. (\ref{eq8}) may
then be expressed in terms of the steady-state covariance matrix $\Lambda
_{\infty }$, i.e. $\langle m_{x}^{i}m_{x}^{j}\rangle =\Lambda _{\infty }%
\left[ 4(i-1)+1,4(j-1)+1\right] $ and $\langle m_{y}^{i}m_{y}^{j}\rangle
=\Lambda _{\infty }\left[ 4(i-1)+2,4(j-1)+2\right] $. 
\begin{figure}[t]
\includegraphics[width=0.5\textwidth]{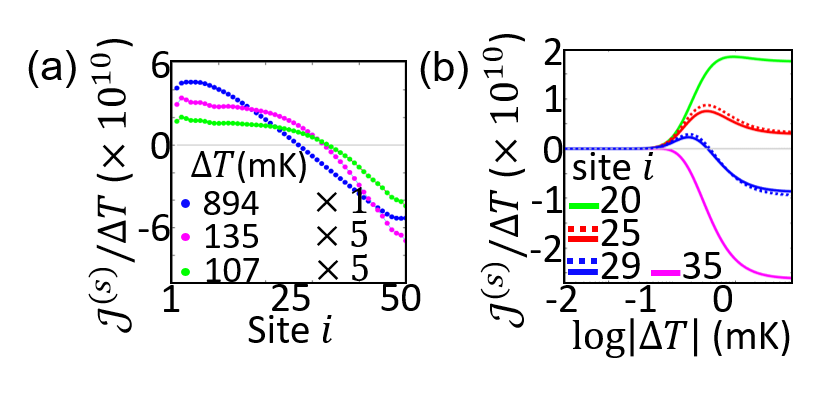} 
\caption{(a) Site dependence of $\mathcal{J}_{i}^{(s)}$ for the $\Delta T$ $%
=107,135,894\,$mK from Fig. \protect\ref{fig2}(c). (b) Dependence of spin
current $\mathcal{J}_{i}^{(s)}\left( \Delta T\right) $ of four sites $i$ on $%
\Delta T$, from Fig. \protect\ref{fig2}(c). The dashed lines correspond to a
finer mesh $d=25\,$nm (rather than 50$\,$nm), illustrating convergence. In
(a) and (b), $T_{L}=20\,$mK.}
\label{fig3}
\end{figure}

\textit{Temperature dependence -} We apply a linear temperature gradient
with $\Delta T=T_{R}-T_{L}$, $T_{L}$ ($T_{R}$) is the temperature at the
left (right) edge of the nanowire. In linear response, the TSSE signal is
antisymmetric, changing sign in the middle of the wire \cite%
{Xiao2010,Adachi2011,Ohe2011,Ritzmann2015,Yan2017}. Figures \ref{fig2}(b)
and (c) show amplitudes (left panels) and signs (right panels) of $\mathcal{J%
}_{i}^{s}/\Delta T$ as a function of $T_{M}=\left( T_{R}+T_{L}\right) /2$
and temperature difference $\Delta T$, respectively, for $w_{y}=500\,$nm and 
$h^{z}=20\,$mT [with dispersion in Fig. \ref{fig2}(a)]. In Fig. \ref{fig2}%
(b), we show the dependence on the mean temperature $T_{M}$ for fixed $%
\Delta T=10\,$mK. In Fig. \ref{fig2}(c), $T_{L}=20\,$mK is fixed and the
gradient $\Delta T$ is varied. According to Figure \ref{fig2}(b), the signal
increases with increasing $T_{M}$. The early saturation is an artifact by
the frequency cut-off, $\omega _{\mathrm{max}}$, introduced by the finite
mesh size $d$, so results are valid for $T<\hbar \omega _{\mathrm{max}%
}/k_{B} $. The qualitative features nevertheless remain intact for half the
mesh size $d$, i.e. a $\sim 4$ times larger cutoff frequency, as illustrated
in Fig. \ref{fig3}(b). The site index $i_{\pm }$ at which $\mathcal{J}%
_{i}^{s}$ changes sign in Fig. \ref{fig2}(b) shifts to the right edge with
decreasing $T_{M}$ below $\hbar \omega _{c}/k_{B}=0.1\,$K, where $\omega
_{c} $ is the uniform (Kittel) mode frequency. According to Fig. \ref{fig2}%
(c), the asymmetry survives at higher temperatures with increasing
temperature difference $\Delta T$ and fixed $T_{L}\ll \hbar \omega
_{c}/k_{B} $. Figures \ref{fig3}(a)-(b) emphasize the essence of the results
in Fig. \ref{fig2}(c) (see also Fig. S1 \cite{SM}). Figure \ref{fig3}(a)
shows the deviation of the signal from an antisymmetric profile. In Fig. \ref%
{fig3}(b), we observe that for a contact on the right half of the nanowire
and fixed small $T_{L}=20\,$mK, the TSSE signal changes sign and a maximum
appears at relatively large $\Delta T$ and $T_{M}$. The dashed curves in
Fig. \ref{fig3}(b) show that for smaller mesh size $d$, i.e. higher cutoff
frequency, the peak and sign change features remain intact. In the blue
curves of Fig. \ref{fig3}(b), the $\Delta T^{\prime }$s that cause the sign
change and peak are $\sim 0.4\,$K and $\sim 0.2\,$K, respectively.

The deviation from an antisymmetric signal can partly be understood in a
semiclassical picture. The additional occupation of a magnon mode with
frequency $\omega _{0}$ scales like $\delta n_{BE}\sim \tau _{r}v(\omega
_{0})\nabla n_{BE}(\omega _{0})$ \cite{Rezende2014}, where $\tau _{r}$ is a
relaxation time, $v({\omega _{0}})$ is the group velocity, and $%
n_{BE}(\omega _{0})=\left( e^{\hbar \omega _{0}/k_{B}T}-1\right) ^{-1}$. $%
\int_{\omega }\delta n_{BE}\neq 0$ because the magnons pile up or get
drained at the edges. In linear response and a long spin-diffusion length
the dependence is linear with a zero in the center. An expansion in $\hbar
\omega _{0}/\left( k_{B}T\right) $ can only indicate that $\nabla
n_{BE}(\omega _{0})$ is uniform for $T_{M}>\hbar \omega _{0}/k_{B}$ when $%
\Delta T\ll T_{M}$ [see Fig. \ref{fig2}(b)], and for $\Delta T\gg \hbar
\omega _{0}/k_{B}$ when $T_{M}\sim \Delta T/2$ [see Fig. \ref{fig2}(c)]. A
detailed calculation is indeed necessary to determine the spatial dependence
of magnon accumulation in the nonlinear regime of temperature, where we
observe a substantial non-monotonicity of signal at certain contact
positions [see Fig. \ref{fig3}(b)].

The TSSE voltage $V_{i}^{(TSSE)}=2\rho \theta _{H}ew_{y}g_{r}\mathcal{J}%
_{i}^{s}/\left( 4\pi w_{y}d\right)$, where $-e$ is the electron charge, and
for Pt, the conductivity $\rho =0.9\,\mathrm{\mu {\Omega }\,}$m \cite%
{Kajiwara2010}, spin Hall angle $\theta _{H}=0.07$ \cite{Liu2011}, while for
the YIG/Pt interface $g_{r}/\left( w_{y}d\right) =10^{16}\,1/\text{m}^{2}$ 
\cite{Kajiwara2010}. This leads to $V_{i}^{(TSSE)}\approx 8\times 10^{-18}%
\mathcal{J}_{i}^{s}\,$V. The low temperature maximum of $\mathcal{J}%
_{i}^{s}/\Delta T\sim 10^{10}$ for $\Delta T=0.2\,$K [see Fig. \ref{fig3}%
(b)] leads to a substantial $V_{i}^{(TSSE)}\sim 16\,$nV.

\begin{figure}[t]
\includegraphics[width=0.5\textwidth]{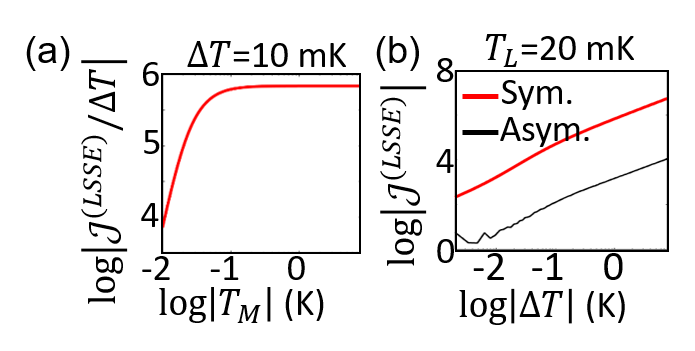}
\caption{Longitudinal spin Seebeck effect. The temperature dependence of the
spin current at the right endpoint flowing into a spin sink (a) $\Delta
T=10\,$mK. (b) $T_{L}=20\,$mK. Here we plot $(\mathcal{J}_{L}^{(LSSE)}-%
\mathcal{J}_{R}^{(LSSE)})/2$ (red line labeled Sym) and $\mathcal{J}%
_{R}^{(LSSE)}+\mathcal{J}_{L}^{(LSSE)}$ (black line labeled Asym.)}
\label{fig4}
\end{figure}

\textit{LSSE -} The LSSE records the total spin current generated in the
magnet within the spin relaxation length and not just the magnon
accumulation at the contact as in the TSSE. We can access the LSSE by
modifying the boundary conditions at the terminals of the wire in order to
allow the spin currents to flow unimpeded into the contacts that act as spin
and energy sinks. To this end, we introduce two reservoirs to the left and
right of the nanowire. We assume the reservoirs to be non-magnetic metals
(NM) with a large spin mixing conductance and interfacial damping at both
ends as detailed in \cite{SM}. This is in contrast to the contacts in the
TSSE, which we assumed to be non-invasive. Figures \ref{fig4}(a) and (b)
show temperature dependence of the average current through the wire $%
\mathcal{J}^{(LSSE)}=\left( \mathcal{J}_{R}^{(LSSE)}-\mathcal{J}%
_{L}^{(LSSE)}\right) /2$. \textbf{\ }The temperature combinations of Figs. %
\ref{fig4}(a) and (b) are the same as in Figs. \ref{fig2}(a) and (b), i.e.
fixed $\Delta T=10\,$mK but varying $T_{M}$ or fixed $T_{L}=20\,$mK but
varying $\Delta T$, respectively. Figures \ref{fig4}(a) and (b) are
featureless and illustrate that even for low $T_{M}\ll \Delta T$, $\mathcal{J%
}^{(LSSE)}$ depends (quasi-)linearly on $\Delta T$ [see Fig. \ref{fig4}%
(b)]. In Fig. \ref{fig4}(b), we also show difference of the spin currents
into the left and out of the right contacts $\mathcal{J}_{R}^{(LSSE)}+%
\mathcal{J}_{L}^{(LSSE)}$, which turns out to be relatively very small
because LSSE is dominated by
the bulk spin current which is the same for both contacts. It should be
emphasized that our TLS model is strictly valid only for $T<1\,$K.

\textit{Conclusion -} We investigate SSE at cryogenic temperatures $%
T\lesssim 1\,$K with dissipation by two-level systems. In the nonlinear
temperature regime, i.e. large $\Delta T/T$, we predict a non-monotonic TSSE
signal at certain position of the detector contacts. For a linear
temperature gradient, and a contact position in the hot region, the sign
changes and a substantially large $\sim 10\,$nV voltage peak emerges at $%
2T_{M}\approx \Delta T\sim 0.2\,$K. On the other hand, the LSSE signal
follows a (quasi-)linear dependence on $\Delta T$, even when much larger
than the average temperature.

\textit{Acknowledgments -}We acknowledge support by JSPS KAKENHI Grants with
Nos. 20H02609 and 19H006450.

\end{document}


\title{Cryogenic spin Seebeck effect: Supporting Material}
\author{Mehrdad Elyasi}
\affiliation{Institute for Materials Research, Tohoku University, 2-1-1 Katahira, 980-8577 Sendai, Japan
}
\author{Gerrit E. W. Bauer}
\affiliation{Institute for Materials Research \& AIMR \& CSRN, Tohoku
University, 2-1-1 Katahira, 980-8577 Sendai, Japan} 
\affiliation{Zernike
Institute for Advanced Materials, University of Groningen, The Netherlands}
\date{\today }
\maketitle

In this supplementary material, we elaborate on the temperature dependence of
the TSSE, explore the effect of the non-monotonicity of the magnon dispersion
caused by dipolar interactions, and explain our model for the terminal
contacts in the LSSE.

\renewcommand{\thefigure}{S1}
\begin{figure}[t]
\includegraphics[width=0.5\textwidth]{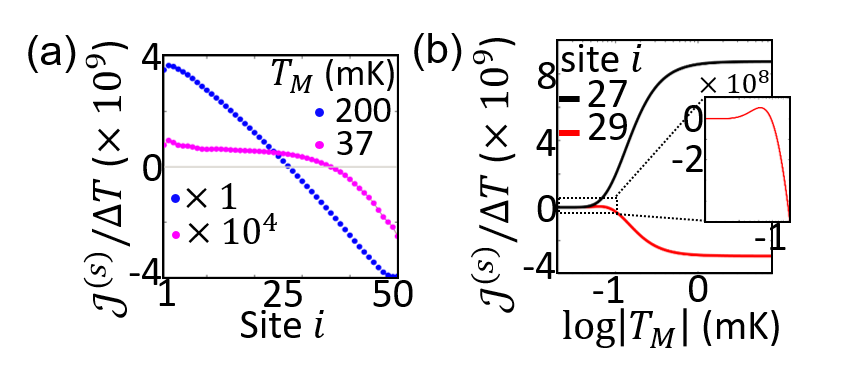}\caption{(a) Site dependence
of $\mathcal{J}_{i}^{(s)}$ for two $T_{M}$. For $T_{M}=37\,$mK, scaled by
$10^{4}$. (b) $T_{M}$ dependence of $\mathcal{J}_{i}^{(s)}$ for two sites $i$.
The inset zooms in on the maximum. In (a) and (b), $\Delta T=10\,$mK. The
plots are cross sections from Fig. 2(b) of the main text.}

\label{fig1}%
\end{figure}

\renewcommand{\thefigure}{S2}
\begin{figure}[t]
\includegraphics[width=0.5\textwidth]{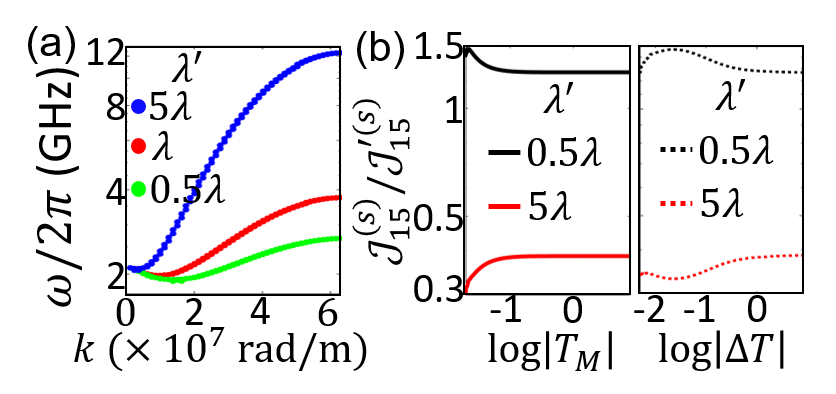}\caption{(a) The magnon
dispersion for three different exchange length values $\lambda^{\prime
}=0.5\lambda$, $\lambda$, and $5\lambda$, where $\lambda$ is that of YIG. (b)
Left (right) panel: $T_{M}$ ($\Delta T$) dependence of the ratio
$\mathcal{J}_{i}^{(s)}/\mathcal{J}_{i}^{\prime(s)}$. $\mathcal{J}_{i}%
^{\prime(s)}$ is TSSE spin current for $\lambda^{\prime}\neq\lambda$. The
results are for the site $i=15$. In (b), vertical axes are logarithmic. }%
\label{fig2}%
\end{figure}

\renewcommand{\thefigure}{S3}
\begin{figure}[t]
\includegraphics[width=0.5\textwidth]{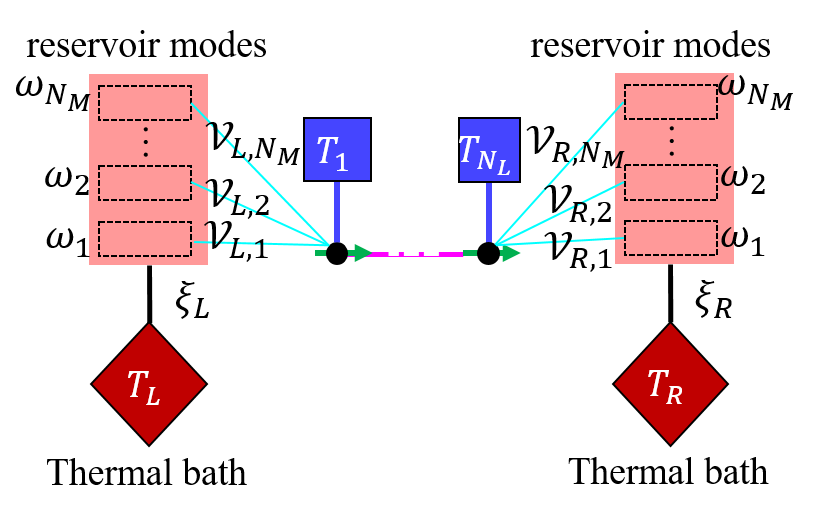}\caption{LSSE. The LSSE
model schematics of additional mesoreservoirs to the left and right edges of
the nanowire [see Fig. 1(c) of the main text]. The mesoreservoirs consist of $N_{M}$
bosonic modes with frequencies $\omega_{q}$, and dissipate ($\xi_{L(R)}$) to a
thermal bath at $T_{L(R)}$.}%
\label{fig3}%
\end{figure}

\textit{TSSE temperature dependence -} Figure \ref{fig1}(a) shows the
deviation of the TSSE current $\mathcal{J}_{i}^{s}$ from an antisymmetric
profile for low $T_{M}$ and $\Delta T=10\,$mK. In Fig. \ref{fig1}(b), we
observe that for contacts on the right half of wire the TSSE signal may change
its sign. Compared to Fig. 3(b) of the main text, this sign change and a local
maximum appear at much larger $\Delta T$ and $T_{M}$, when $T_{L}$ (rather
than $T_{M})$ is fixed and small. Here we estimate a TSSE voltage
$V_{i}^{(TSSE)}\approx8\times10^{-18}\mathcal{J}_{i}^{s}\,$V. The low
temperature maximum in Fig. \ref{fig1}(b), $\mathcal{J}_{i}^{s}/\Delta
T\sim10^{8}$ at\textbf{\ }$\Delta T=10\,$mK, generates only a very small
$V_{i}^{(TSSE)}\sim8\,$pV.

\textit{Dispersion monotonicity effect -} Here we discuss a consequence of the
magnetodipolar interaction that generates negative group velocities at the
origin and a mininum of the magnon dispersion at finite wave numbers. We can
study the latter as a function of decreasing $w_{y},$ which renders the
minimum to become shallower and finally vanishes. In order to explore the
effects of non-monotonous dispersion monotonicity we
can as well change the exchange length $\lambda$ for fixed $w_{y},$ which is
not physical but sufficient to illustrate that the effect is small. Figure
\ref{fig2}(a) shows the dispersion for exchange lengths $\lambda^{\prime}%
\in\{0.5,1,5\}\lambda$, where $\lambda$ is that of YIG. The dispersion is
deeper (shallower) for smaller (larger) $\lambda^{\prime}$ as expected. Figure
\ref{fig2}(b) shows the temperature dependence of $\mathcal{J}_{i}%
^{(s)}/\mathcal{J}_{i}^{\prime(s)}$ at a representative site $i=15$, where
$\mathcal{J}_{i}^{\prime(s)}$ is the TSSE current corresponding to
$\lambda^{\prime}$. For larger $\lambda^{\prime}$, i.e. a shallower valley,
the TSSE signal increases with the absolute value of the average group
velocity over the occupied states. For magnetization $\Vert\hat{x}(\hat{y})$,
the dispersion increases monotonically which increases the spin currents, but
detection by the ISHE becomes more complicated (not shown).

\textit{LSSE model-} The LSSE records the total spin current generated in the
magnet within the spin relaxation length and not just the magnon accumulation
at the contact as in the TSSE. We can access the LSSE by modifying the
boundary conditions at the terminals of the wire in order to allow the spin
currents to flow unimpeded into the contacts that act as spin and energy
sinks. To this end, we introduce two reservoirs to the left and right of the
nanowire. We assume the reservoirs to be non-magnetic metals (NM) with a large
spin mixing conductance and interfacial damping at both ends, in contrast to
the contacts in the TSSE, which we assumed to be non-invasive. We model the
end contacts by reservoirs coupled equally to all modes in the nanowire by a
structured reservoir model
\cite{Leggett1987,Carmichael1999,Prosen2010,Ajisaka2012} of $N_{M}$ bosonic
modes, as depicted in Fig. \ref{fig3}(a). The modified total Hamiltonian
$H^{\prime}=H+H_{SR}^{\prime}$, where $H_{SR}^{\prime}=\sum_{q}\left(
\mathcal{V}_{L,q}e_{L,q}^{\dag}a_{1}+\mathcal{V}_{R,q}e_{R,q}^{\dag}a_{N_{L}%
}+\mathrm{H.c}.\right)  $, $e_{L(R),q}^{\dag}$ creates a boson in
mode $q$ of the left (right) reservoir, and $\mathcal{V}_{L(R),q}$ is the
coupling of the local magnon field of the left (right) edge to the left
(right) reservoir. Each boson mode in the left (right) reservoir is in contact
with a large thermal bath at $T_{L}$ ($T_{R}$), and dissipation $\xi_{L}$
($\xi_{R}$). This dissipation is accompanied by the fluctuating field
$g_{L(R),q}^{\prime}=\sqrt{\xi_{L(R)}}\mathcal{F}_{L(R),q}^{\prime}$, where
$\langle\mathcal{F}_{L(R),q}^{\prime}(t)\mathcal{F}_{L(R),q}^{\prime\dag
}(t^{\prime})\rangle=\left(  n_{L(R),q}^{th}+1\right)  \delta\left(
t-t^{\prime}\right)  $, $\langle\mathcal{F}_{L(R),q}^{\prime\dag
}(t)\mathcal{F}_{L(R),q}^{\prime}(t^{\prime})\rangle=n_{L(R),q}^{th}%
\delta\left(  t-t^{\prime}\right)  $, $n_{L(R),q}^{th}=\left(  e^{\hbar
\omega_{q}/k_{B}T_{L(R)}}-1\right)  ^{-1}$, assuming $1/\xi_{L(R)}\gg
\hbar/k_{B}T_{L(R)}$. The density of states of the left (right) reservoir
$\mathcal{G}_{L(R)}(\omega)=\sum_{q}2\xi_{L(R)}\omega^{2}\left[  \left(
\omega_{q}^{2}-\omega^{2}\right)  ^{2}+\omega^{2}\xi_{L(R)}^{2}\right]  ^{-1}%
$. Therefore, the dissipation of a magnon with frequency $\omega_{q}$ in the
left (right) magnetic site to the left (right) reservoir $\Gamma
_{L(R)}^{\prime}(\omega_{q})=\mathcal{G}_{L(R)}(\omega_{q})\mathcal{V}%
_{L(R),q}^{2}$. The reservoirs should dissipate magnons of all frequencies, so
we impose $N_{M}=N_{L}$ and $\omega_{q}$ to be frequencies of the normal
magnon modes of the nanowire. The spin and energy loss at the edges is
equivalent to an increased damping $\alpha_{SP}$, i.e. $\Gamma_{L(R)}^{\prime
}(\omega_{q})=\alpha_{SP}\omega_{q}$, so $\mathcal{V}_{L(R),q}=\sqrt
{\alpha_{SP}\omega_{q}/\mathcal{G}(\omega_{q})}$. For $2\pi/\xi_{L(R)}=10\,$ns
and $\alpha_{SP}=0.1\alpha_{TLS}$, we obtain the steady state covariance matrix
$\Lambda_{\infty}^{\prime}$ as described in the main text, that allows computing the
spin current flowing from the left (right) reservoir into (out of) the
nanowire $\mathcal{J}_{L(R)}^{(LSSE)}=-i\left[  S_{1(N_{L})}^{z},H\right]  $
as
\begin{align}
\mathcal{J}_{L(R)}^{(LSSE)} &  =\frac{1}{2}\sum_{j}\left\{  \left[
SJ\delta\left(  1(N_{L})\pm1,j\right)  +\mathcal{B}_{1(N_{L}),j}\right]
\times\right. \nonumber\\
&  \left(  \langle m_{x}^{1(N_{L})}m_{y}^{j}\rangle-\langle m_{y}^{1(N_{L}%
)}m_{x}^{j}\rangle\right)  +\nonumber\\
&  \left.  \mathcal{C}_{1(N_{L}),j}\left(  -\langle m_{x}^{1(N_{L})}m_{y}%
^{j}\rangle-\langle m_{y}^{1(N_{L})}m_{x}^{j}\rangle\right)  \right\}
.\label{eq7}%
\end{align}